\documentclass{article}
\usepackage{epsfig}
\usepackage{amsmath,amssymb}

\usepackage{latexsym,graphicx,epsfig}
\usepackage{curves,eepic}
\def\lb{\linebreak[4]}
\newcommand{\be}{\begin{equation}}
\newcommand{\ee}{\end{equation}}
\newcommand{\bes}{\begin{subequations}}
\newcommand{\ees}{\end{subequations}}
\newcommand{\bea}{\begin{eqnarray}}
\newcommand{\eea}{\end{eqnarray}}
\newcommand{\bear}{\begin{equation}\begin{array}}
\newcommand{\eear}[1]{\end{array}\label{#1}\end{equation}}
\def\ba{$$\begin{array}}
\def\ea{\end{array}$$}
\def\bra{$\begin{array}}
 \def\era{\end{array}$}

\newcommand{\bm}{\boldmath}
\newcommand{\fr}[2]{\dfrac{{ #1}}{{ #2}}}

\newcommand{\pa}{\partial}
\newcommand{\la}{\langle}
\newcommand{\ra}{\rangle}

\def\vak{{\varkappa}}

\def\cl{\centerline}

\newsavebox{\fmbox}

{\end{list}}
\newcounter{enumct}

\newcommand{\bu}{$\bullet$\ }

\begin{document}
\title{Symmetries of 2HDM, different vacua,\\
CP violation\\ and  possible relation to a history of time
}

\author{{\it I. F. Ginzburg,}\\
{ Sobolev Institute of Mathematics,  Novosibirsk, Russia}}

\date{}
\maketitle

\cl{\it To be published in Proc. Photon05, Acta Physica
Polonica}

\begin{abstract}
The same physical reality in Two Higgs doublet model (2HDM)
can be described by different Lagrangians. We call this
property the reparametrization invariance (in space of
Lagrangians) and study corresponding symmetry group and its
subgroup describing rephasing invariance.  Next we consider
the $Z_2$-symmetry of the Lagrangian, which prevents a
$\phi_1\leftrightarrow\phi_2$ transitions, and  the
different levels of its violation, soft and hard.  We argue
that the 2HDM with a soft breaking of $Z_2$-symmetry is a
natural model in the description of EWSB. We also consider
vacuum structure of the 2HDM. We find very simple condition
for a CP violation in the Higgs sector. In the Model~II for
Yukawa interactions we obtain the set of relations among
the couplings to gauge bosons and to fermions which allows
one to analyze different physical situations (including CP
violation) in terms of these very couplings, instead of the
parameters of Lagrangian.\\
We discuss possible interaction of Higgs fields of SM or
2HDM with inflatory Higgs field describing exponential
expansion of Universe after Big Bang and possible variation
in the scenario of beginning of Time.
\end{abstract}

\maketitle

\section{Lagrangian}


A spontaneous electroweak symmetry breaking of  (EWSB) via
the Higgs mechanism is described by the Lagrangian
 \begin{subequations}\label{Higgslagr}\\[-4mm]
 \begin{equation}
{ \cal L}={ \cal L}^{SM}_{ gf } +{ \cal L}_H + {\cal L}_Y
\;\; \;with\;\;\; { \cal
L}_H=T-V\,.\label{Eq:Lagr-Higgs}\vspace{-2mm}
 \end{equation}
Here ${\cal L}^{SM}_{gf}$ describes the $SU(2)\times U(1)$
Standard Model interaction of gauge bosons and fermions,
${\cal L}_Y$ describes the Yukawa interactions of fermions
with Higgs scalars and ${\cal L}_H$ is the Higgs scalar
Lagrangian; $T$ is the Higgs kinetic term and $V$ is the
Higgs potential.

In the minimal Standard Model (SM) one scalar isodoublet
with hypercharge $Y=1$ is implemented. Here ${\cal
L}_H=(D_\mu\phi)^\dagger D_\mu\phi-V$, with  the Higgs
potential $V=\lambda\phi^4/2-m^2\phi^2/2$ etc. In
ref.~\cite{GK05} we study in detail the simplest extension
of the SM (see \cite{Hunter} for earlier references), with
two scalar fields $\phi_i$ being weak isodoublets ($T=1/2$)
with hypercharges $Y= 1$ called the Two-Higgs-Doublet Model
(2HDM). The kinetic term of the most general renormalizable
Higgs Lagrangian is\\[-3mm]
 \be T= (D_{\mu}
\phi_1 )^{ \dagger}(D^{\mu} \phi_1)+ (D_{\mu} \phi_2 )^{
\dagger}(D^{\mu} \phi_2) +\left[\vak (D_{\mu} \phi_1 )^{
\dagger}(D^{\mu} \phi_2) +h.c.\right]
\label{kinterm}\vspace{-3mm}
 \ee
and the Higgs potential, containing operators of dimension
2 (in mass term) \eqref{massL} and of dimension 4
\eqref{baspot}, is\\[-7mm]
 \bea
 & V=-\dfrac{1}{2}\left\{m_{11}^2(\phi_1^\dagger\phi_1)+
 m_{22}^2(\phi_2^\dagger\phi_2)
+\left[m_{12}^2 (\phi_1^\dagger\phi_2)+{\rm h.c.}\right]
\right\}\label{massL}\\
&\begin{array}{c}
+\dfrac{\lambda_1}{2}(\phi_1^\dagger\phi_1)^2
+\dfrac{\lambda_2}{ 2}(\phi_2^\dagger\phi_2)^2
+\lambda_3(\phi_1^\dagger\phi_1) (\phi_2^\dagger\phi_2)
+\lambda_4(\phi_1^\dagger\phi_2)
(\phi_2^\dagger\phi_1)\\[2mm]
   +\dfrac{1}{2}\left[\lambda_5(\phi_1^\dagger\phi_2)^2+{\rm
h.c.}\right]
+\left\{\left[\lambda_6(\phi_1^\dagger\phi_1)+\lambda_7
(\phi_2^\dagger\phi_2)\right](\phi_1^\dagger\phi_2) +{\rm
h.c.}\right\} \end{array}\label{baspot}
 \eea
  \end{subequations}

\section{Reparametrization and rephasing
invariance.}

\subsection{Reparametrization (RPa) invariance.}

Our model contains  two fields with identical quantum
numbers. Therefore,  it can be described in similar way
both in terms of fields $\phi_k$ $(k=1,2)$,  used in
(\ref{Higgslagr}), and in terms of fields $\phi'_k$
obtained from $\phi_k$ by a global unitary transformation
${\cal {\hat F}}$ of $SU(2)\times U(1)$ {\it general
reparametrization (RPa) group}:\\[-3mm]
 \be
\begin{pmatrix}\phi_1'\\
\phi_2'\end{pmatrix}=\hat{\cal F}\begin{pmatrix}\phi_1\\
\phi_2\end{pmatrix}\,,\quad
 \hat{\cal F}=e^{-i\rho_0}\begin{pmatrix}
\cos\theta\,e^{i\rho/2}&\sin\theta\,e^{i(\tau-\rho/2)}\\
-\sin\theta\,e^{-i(\tau-\rho/2)}&\cos\theta\,e^{-i\rho/2}
\end{pmatrix}.
\label{reparam}\vspace{-3mm}
 \ee
This group splits into proper  $SU(2)$ RPa group with
parameters $\theta$, $\rho$, $\tau$ ({\it similar to the
gauge parameter of gauge theories}) and $U(1)$ group
describing {\it overall phase freedom}, with parameter
$\rho_0$.

\bu This transformation induces the changes of coefficients
of Lagrangian ({\it independent on parameter  $\rho_0$}),
$\lambda_i\to\lambda'_i$ and $m_{ij}^2\to (m')_{ij}^2$,
$\vak\to \vak'$ with renormalization of fields
$\phi_i^\prime$  (RPa transformation of parameters). They
are presented at $\vak=0$ in ref.~\cite{GK05}. By
construction, the Lagrangian of the form (\ref{Higgslagr})
with coefficients
$\lambda_i$, $m_{ij}^2$ and that with 
new coefficients $\lambda'_i$, ${(m')}_{ij}^2$ describe the same
physical reality.  We call this property a {\it RPa invariance in  a
space of Lagrangians} (with coordinates given by its parameters).

The set of RPa transformations for parameters of Lagrangian
form representation of RPa group in the {\it 16-dimensional
space of Lagrangians} with coordinates given by
$\lambda_{1-4}$, $Re\lambda_{5-7}$, $Im\lambda_{5-7}$,
$m^2_{11,22}$, $Re(m^2_{12})$, $Im(m^2_{12})$, $Re\vak$,
$Im\vak\,$.  A set of physically equivalent Higgs
Lagrangians, obtained from each other by mentioned
transformations, forms {\it the reparametrization
equivalent space} (RPaES), being a 3-dimensional subspace
of the entire space of Lagrangians. The parameters of
Lagrangian can be determined from measurements in principle
only with accuracy up to the RPa freedom.

All observable quantities are invariants of RPa
transformations \mbox{(IRpaT)}. These are, for example,
masses of observable Higgs bosons -- eigenvalues of mass
matrix \eqref{Eq:M3by3} and \eqref{Eq:mch} and eigenvalues
of Higgs-Higgs scattering matrices \eqref{scmat}. The
coefficients of secular equation for diagonalization of
these matrices (among them -- trace of this matrix and its
determinant) can be constructed from these eigenvalues.
Therefore, they are also IRpaT's.

By writing Higgs potential as a sum
$Y_{ab}(\phi^\dagger_a)\phi_b)
+Z_{abcd}(\phi^\dagger_a)\phi_b)(\phi^\dagger_c)\phi_d)$
with $a, \, b=1,\,2$, one can construct IRPaT's  at
$\vak=0$ as combinations of products of $Y$ and $Z$
summarized on $a,\,b$. In this way large series of
(generally not independent) IRPaT's was obtained
\cite{GH05}. The group--theoretical analysis of RPa group
allows one to find the {\bf complete} set of {\it
independent} IRPaT (7 constructed from $\lambda$'s and a
few more with $m_{ij}$, $\vak$)~\cite{Iv05}.

\subsection{Rephasing (RPh) invariance}

It is useful  to consider a particular case of the
transformations (\ref{reparam}) with $\theta=0$ -- global
{\em rephasing (RPh) transformation   of the fields}:\\[-3mm]
 \bes\label{Eq:rephase}
 \be
\phi_k\to e^{-i\rho_i}\phi_k,\qquad
\rho_1=\rho_0-\rho/2,\;\,\rho_2=\rho_0+\rho/2,\;\,
\rho=\rho_2-\rho_1.\label{rephase}\vspace{-2mm}
 \end{equation}
This transformation leads to  {\it a RPh transformation of
the  Lagrangian}:\\[-3mm]
 \begin{equation}
\begin{array}{c}
\lambda_{1-4}\to \lambda_{1-4},\;\,
m_{11}^2\to m_{11}^2,\;\, m_{22}^2\to m_{22}^2,\\
\lambda_5\to\lambda_5\, e^{-2i\rho},\;\;
\lambda_{6,7}\to\lambda_{6,7}\,e^{-i\rho},\;\;m_{12}^2\to
m_{12}^2e^{-i\rho}, \,\;\vak\to
\vak\,e^{-i\rho}.\vspace{-2mm}
\end{array}\label{Eq:rephaselam}
\end{equation}\ees

By construction, the Lagrangian of the form
(\ref{Higgslagr}) with coefficients $\lambda_i$, $m_{ij}^2$
and  that with coefficients given by
eq.~(\ref{Eq:rephaselam}) describe the same physical
reality. We call this property a {\it RPh invariance}. The
transformations (\ref{Eq:rephase}) represent $U(1)$ {\it
RPh transformation group} with parameter $\rho$. This RPh
group is a subgroup of the $SU(2)$ RPa group.

\section{Lagrangian and  $Z_2$ symmetry}

One of the earliest reasons for introducing the 2HDM was to
describe the phenomenon of CP violation \cite{Lee:1973iz}.
The CP violation and the flavour changing neutral currents
(FCNC) can be naturally suppressed by imposing on the
Lagrangian a $Z_2$ symmetry \cite{Glashow:1977nt}, inhibit
the $\phi_1\leftrightarrow\phi_2$ transitions, that is the
invariance on the Lagrangian under the interchange\\[-4mm]
  \begin{equation}        \label{Eq:Z2-symmetry}
\phi_1 \leftrightarrow\phi_1,\phi_2 \leftrightarrow-\phi_2
\; \text{or} \; \phi_1 \leftrightarrow-\phi_1,
\phi_2\leftrightarrow\phi_2.\vspace{-2mm}
 \end{equation}

\bu {\bf The case of exact \bm{$Z_2$} symmetry} is
described by the Lagrangian ${\cal L}_H$ (\ref{Higgslagr})
with $\lambda_6=\lambda_7=\vak= m_{12}^2 =0$ and only one
parameter $\lambda_5$ can be complex. The RPh
transformation (\ref{Eq:rephase}) with a suitable phase
$\rho$ allows one to get  Lagrangian with a real
$\lambda_5$, within  the RPh invariant space.

\bu {\bf In case of soft violation of $Z_2$ symmetry} one
adds to the $Z_2$ symmetric Lagrangian the term of operator
dimension 2,  $m_{12}^2 (\phi_1^\dagger\phi_2) + h.c. $,
with a generally complex $m_{12}^2$ (and $\lambda_5$)
parameter. This type of violation respects the $Z_2$
symmetry at small distances (much smaller than $1/M$) in
all orders of perturbative series, i.e. the amplitudes for
$\phi_1\leftrightarrow\phi_2$ transitions disappear at
virtuality $k^2 \sim M^2 \to\infty$. That's why we call it
a "soft" violation. The  RPh transformations
\eqref{Eq:rephase} applied to the Lagrangian with a soft
violation of $Z_2$  generate a whole {\it  soft $Z_2$
violating Lagrangian family}.

The general RPa transformation converts the Lagrangian with
exact or softly violated $Z_2$ symmetry ${\cal L}_s$ to a
{\em hidden soft $Z_2$ violation} form ${\cal L}_{hs}$ with
$\lambda_6,\,\lambda_7\neq 0$, $\vak=0$. 14 parameters of
${\cal L}_{hs}$ are constrained since they can be obtained
from 9 independent parameters of an initial Lagrangian
${\cal L}_s$ (+ 3 RPa group parameters), nondiagonal $\vak$
kinetic term don't appear from loop corrections. For such
physical system {\it ${\cal L}_s$ is preferable RPa
representation.}

\bu  In  general case the terms of the operator dimension
4,  with generally complex parameters $\lambda_6$,
$\lambda_7$ and $\vak$, are added to the Lagrangian with a
softly violated $Z_2$ symmetry. In the case of {\bf\bm the
true hard violation of $Z_2$ symmetry} this Lagrangian
cannot be transformed to the exact or softly violated $Z_2$
symmetry form by any RPa transformation, the $Z_2$ symmetry
is broken {\it at both large and small distances} in any
scalar basis.

The mixed kinetic terms (\ref{kinterm}) can be eliminated
by the nonunitary transformation
(rotation + renormalization), e.g.\\[-4mm]
 \be\label{diagkap}
\!\!(\phi_1^{\,\prime},\!\phi_2^{\,\prime})\!\to\!\!
\left(\!\dfrac{\sqrt{\vak^*}\phi_1\!+\!\sqrt{\vak}\phi_2}
{2\sqrt{|\vak|(1\!+\!|\vak|)}}\!\pm\!
\dfrac{\sqrt{\vak^*}\phi_1\!-\!\sqrt{\vak}\phi_2}{2\sqrt{|\vak|(1\!-\!|\vak|)}}
\!\right)\!.\!\!\vspace{-3mm}
 \ee
However, in  presence of the $\lambda_6$ and $\lambda_7$
terms, the renormalization of quadratically divergent,
non-diagonal two-point functions leads anyway to  the mixed
kinetic terms (e.g. from loops with
$\lambda_6^*\lambda_{1,3-5}$ and
$\lambda_7^*\lambda_{2-5}$). It means that  $\vak$\ \
becomes nonzero at the higher orders of perturbative theory
(and {\it vice versa} a  mixed kinetic term generates
counter-terms with $\lambda_{6,7}$). Therefore all of these
terms should be included in Lagrangian
(\ref{Eq:Lagr-Higgs}) on the same footing, i.e. the
treatment of the true hard violation of $Z_2$ symmetry
without $\vak$\ terms is inconsistent. The  parameter
$\vak$ is running like  $\lambda$'s. (This term does not
appear if parameters $\lambda_i$ are constrained by
relations of hidden soft violation of $Z_2$ symmetry.)
Therefore, the diagonalization \eqref{diagkap} is scale
dependent, and the Lagrangian {\it remains off--diagonal}
in fields $\phi_{1,2}$ even at very small distances in any
RPa representation. Such theory seems to be {\it
unnatural}.

Although in  \cite{GK05} and in this paper  we present
relations for the case of hard violation of $Z_2$ symmetry
at $\vak=0$, the  loop corrections can change results
significantly. {\it Such treatment of the case with true
hard violation of $Z_2$ symmetry is as incomplete as in
most  of the  papers considering this "most general 2HDM
potential".}

\section{Vacua}

The extremes of the potential define the  vacuum
expectation values (v.e.v.'s) $\la\phi_{1,2}\ra$ of the
fields $\phi_{1,2}$ via equations:\\[-4mm]
\begin{equation}          \label{Eq:min-cond}
\partial V/\partial\phi_i
|_{\phi_i=\langle\phi_i\rangle} =0.\vspace{-2mm}
\end{equation}
These equations have trivial electroweak symmetry conserving
solution\lb $\la\phi_i\ra=0$ and electroweak symmetry violating
solutions, discussed below. With accuracy to the choice of $z$ axis
in the weak isospin space, and using the overall phase freedom of
the Lagrangian to choose one v.e.v. real, in the 
minimal SM such equation has single EWSB solution\\[-3mm]
 \be
\langle\phi\rangle =\fr{1}{\sqrt{2}}\left(\begin{array}{c} 0\\
v\end{array}\right),\quad v=m/\sqrt{2\lambda}\,.
\label{SMvac}\vspace{-2mm}
 \ee
With the same accuracy the most general electroweak
symmetry violating solution of \eqref{Eq:min-cond} can be
written in a
form\\[-4mm]
 \be
\langle\phi_1\rangle =\fr{1}{\sqrt{2}}\left(\begin{array}{c} 0\\
v_1\end{array}\right),\quad \;\; \langle\phi_2\rangle
=\fr{1}{\sqrt{2}}\left(\begin{array}{c}u \\ v_2
 e^{i\xi}\end{array}\right).\label{genvac}\vspace{-2mm}
 \ee

To describe these extremes it is useful to denote\\[-3mm]
 \bear{c}
x_1=(\phi_1^\dagger\phi_1),\;\;
x_2=(\phi_2^\dagger\phi_2),\;\;
x_3=(\phi_1^\dagger\phi_2),\\ y_1=\la x_1\ra,\;\; y_2=\la
x_2\ra,\;\;y_3=\la x_3\ra, \quad
Z=y_3^*y_3-y_1y_2.\vspace{-1mm}
 \eear{yidef}

It is easy to check that $\pa x_1/\pa\phi_2=\pa
x_2/\pa\phi_1=0$ and\\[-3mm]
 $$\begin{array}{c}
 x_3\left(\fr{\pa x_1}{\pa\phi_1}\phi_1\right)-
 x_1\left(\fr{\pa x_1}{\pa\phi_1}\phi_2\right)=
x_3^*\left(\fr{\pa x_3^*}{\pa\phi_1}\phi_2\right)-
x_2\left(\fr{\pa x_3^*}{\pa\phi_1}\phi_1\right)
 =0,\\[3mm]
x_3\!\left(\fr{\pa
x_3^*}{\pa\phi_1}\phi_1\!\!\right)\!\!-\!
x_1\!\left(\fr{\pa x_3^*}{\pa\phi_1}\phi_2\!\!\right) \!=\!
x_3^*\!\left(\fr{\pa
x_1}{\pa\phi_1}\phi_2\!\!\right)\!\!-\! x_2\!\left(\fr{\pa
x_1}{\pa\phi_1}\phi_1\!\!\right)
 \!=\! x_3x_3^*\!-\!x_1x_2.\vspace{-2mm}
\end{array}$$

Now  the extremum condition \eqref{Eq:min-cond} can be
rewritten as\\[-3mm]
 \bear{c}
\left\la\! x_3 \!\left(\!\fr{\pa V}{\pa
\phi_1}\phi_1\!\!\right)\! -\!  x_1\! \left(\!\fr{\pa
V}{\pa \phi_1}\phi_2\!\!\right)\!\!\right\ra\!=\!Z\!
\left(\!\!\lambda_4y_3\!+\! \!\lambda_5^*\!
y_3^*\!+\!\lambda_6^* y_1\!\!+\!\lambda_7^*
y_2\!-\!\fr{m_{12}^{* 2}}{2}\!\right)\!\!=\!0,\\
\left\la\! x_3^*\!\! \left(\!\fr{\pa V}{\pa
\phi_1}\phi_2\!\!\right)\! -\! x_2 \!\left(\!\fr{\pa V}{\pa
\phi_1}\phi_1\!\!\right)\!\!\right\ra\!=\!Z\!\left(\!\!\lambda_1
y_1\!+\! \lambda_3y_2\!+\!\lambda_6^*
y_3^*\!\!+\!\lambda_6y_3\!
\!-\!\fr{m_{11}^2}{2}\!\!\right)\!\!=\!0,\\
\left\la\! x_3\! \left(\!\fr{\pa V}{\pa
\phi_2}\phi_1\!\!\right)\! -\!  x_1\! \left(\!\fr{\pa
V}{\pa
\phi_2}\phi_2\!\!\right)\!\!\right\ra\!=\!Z\!\left(\!
\lambda_2y_2\!\!+\! \lambda_3y_1\!\!+\!\lambda_7^*
y_3^*\!\!+\!\lambda_7
y_3\!-\!\fr{m_{22}^2}{2}\!\!\right)\!\!=\!0.\vspace{-2mm}
 \eear{nearfin}

Therefore, two opportunities can be realized, in dependence
of zero or nonzero value of $ Z=y_3^*y_3-y_1y_2$. Depending
on the  parameters of potential,  these solutions describe
either saddle point or a minimum of the potential. The
condition for minimum is that all eigenvalues of Higgs mass
matrix are positive, and vacuum energy of one of these
states is smaller than of second.

\subsection{$u\neq 0$ solution, charged vacuum}

We denote by {\it charged vacuum} solution appeared at
 \be
Z=y_3^*y_3-y_1y_2\neq 0\;\;\Rightarrow\;\; u\neq
0\,.\vspace{-2mm}
 \label{chvac}\ee
In this case the v.e.v.'s are given by equations followed
directly from \eqref{nearfin}   \\[-3mm]
  \bear{c}
\lambda_1 y_1\!+\! \lambda_3y_2\!+\!\lambda_6^*
y_3^*+\lambda_6y_3=m_{11}^2/2,\\[1mm]
\lambda_2y_2\!+\! \lambda_3y_1\!+\!\lambda_7^*
y_3^*+\lambda_7 y_3=m_{22}^2/2,\\[1mm]
\lambda_4y_3^*\!+\! \lambda_5 y_3\!+\!\lambda_6
y_1+\lambda_7 y_2=m_{12}^2/2.\vspace{-2mm}
 \eear{chargevac}
With these $y_i$ the Higgs potential \eqref{Higgslagr} can
be written via $\bar{x}_i=x_i-y_i$ as ($E_{vac}^{c}$ is a
vacuum energy)\\[-3mm]
 \bear{c}
V=\lambda_1\bar{x}_1^2/2
+\lambda_2\bar{x}_2^2/2+\lambda_3\bar{x}_1\bar{x}_2+
\lambda_4\bar{x}_3^*\bar{x}_3\\
+\left[\lambda_5\bar{x}_3^2/2+
(\lambda_6\bar{x}_1+\lambda_7\bar{x}_2)\bar{x}_3+h.c.\right]
+E_{vac}^{c}.\vspace{-2mm}
 \eear{potcharg}

In this case it is not  possible to split the gauge boson
mass matrix into a neutral and charged sector,  the
interaction of gauge bosons with fermions will not conserve
electric charge,  photon become massive, etc.
\cite{Diaz-Cruz:1992uw}.  Certainly, this case does not
realized in our World.

\subsection{$u=0$ solution, physical (neutral)
vacuum}

We consider below (except final section) only solution of
extremum condition \eqref{Eq:min-cond}, obeying a condition
for $U(1)$ symmetry of electromagnetism,\\[-3mm]
 \be
Z=  y_3^*y_3-y_1y_2=0\,\,\Rightarrow \langle\phi_1\rangle\!
=\!\dfrac{1}{\sqrt{2}}
\begin{pmatrix}
0\\ v_1
\end{pmatrix},\, \,
\langle\phi_2\rangle
\!=\!\dfrac{1}{\sqrt{2}}\begin{pmatrix}0 \\
v_2 e^{i\xi}\end{pmatrix}.
  \vspace{-2mm} \label{phvac2}
  \ee
The other standard notations are $v_1=v\cos\beta$,
$v_2=v\sin\beta$, with SM constraint $v=(\sqrt{2}G_{\rm
F})^{-1/2}=246\,GeV$.

The rephasing of fields  (\ref{rephase})  shifts the phase
difference $\xi$ as $ \xi\to\xi-\rho$.

Let us take some Lagrangian describing our model and
calculate v.e.v.'s. Than, by making the RPh transformation
with $\rho=\xi$, we get {\em the real vacuum Lagrangian}
with real $v_2$ and with parameters, supplied
for a moment by subscript {\it rv}:\\[-3mm]
 \bear{c}\label{newlam}
\lambda_{1-4,rv}=\lambda_{1-4},\;\,
\lambda_{5,rv}=\lambda_5e^{-2i\xi},\;\,
\lambda_{6,rv}=\lambda_{6}e^{-i\xi}, \;\,
\lambda_{7,rv}=\lambda_{7}e^{-i\xi},\\[2mm]  \vak_{rv} = \vak
e^{-i\xi}, \;\, m_{12,rv}^2=
m_{12}^2e^{-i\xi}.\vspace{-1mm}\end{array}
 \end{equation}
The set of real vacuum Lagrangians forms a subspace in the
entire RPaES. In different points of this subspace the
$\tan\beta$ values are different.

The following combinations of parameters and new quantities
are useful:\\[-3mm]
  \bear{c}
\lambda_{3,rv}+\lambda_{4,rv}+Re\lambda_{5,rv}=\lambda_{345,rv}\,,\;\;
 \dfrac{v_1}{v_2}\lambda_{6,rv}\pm
 \dfrac{v_2}{v_1}\lambda_{7,rv}=
\lambda_{67,rv}^\pm\,,\\[2mm]
m_{12,rv}^2=2v_1v_2(\nu+i\delta).\vspace{-2mm}\end{array}\label{nudeldef}
 \ee

The minimum conditions \eqref{yidef}   for this form of
Lagrangian are written as\\[-3mm]
 \bear{c}
(m_{11}^2\!-\!Re\,
m_{12}^2v_2/v_1)\!/2\!-\!\lambda_1v_1^2\!+\!\lambda_{345}v_2^2\!+\!
Re\left(3\lambda_6
v_1v_2\!+\!\lambda_7v_2^3/v_1\right)\!=\!0,
 \\[1mm]
(m_{22}^2\!-\!Re\,
m_{12}^2v_2/v_1)\!/2\!-\!\lambda_2v_2^2\!+\!\lambda_{345}v_1^2\!+\!
Re\left(3\lambda_7
v_1v_2\!+\!\lambda_6v_1^3/v_2\right)\!=\!0,
 \\[1mm]
Im\, m_{12}^2\equiv 2v_1v_2\delta= v_1v_2
Im\,\left(\lambda_{5}
+\lambda_{67}^+\right)\,.\vspace{-1mm}
 \eear{newlamconstr}

We prepare calculations below for  real vacuum potential,
describing it in terms of $v_1$, $v_2$ instead of three
quadratic parameters $m_{11,22}^2, m_{12}^2$
\eqref{newlamconstr}. In this way $\nu\propto Re\,
m_{12,rv}^2$ is {\it single free} parameter in addition to
$v_{1,2}$ while $\delta\propto Im\, m_{12,rv}^2$ is
expressed via $Im(\lambda_{5-7,rv})$ \eqref{newlamconstr}.

\section{Physical Higgs representation }

A standard decomposition of the fields $\phi_{1,2}$ in the
component fields is\\[-3mm]
\begin{equation}
\phi_1\!=\!\begin{pmatrix} \varphi_1^+ \\
(v_1+\eta_1+i\chi_1)/\sqrt{2}
\end{pmatrix}, \;\,
\phi_2=\begin{pmatrix} \varphi_2^+ \\
(v_2+\eta_2+i\chi_2)/\sqrt{2}
\end{pmatrix}.\label{videf}\vspace{-1mm}
\end{equation}
At $\vak=0$ such decomposition conserve a diagonal form of
kinetic terms for fields $\varphi_i^+,\,\chi_i,\,\eta_i$.
The  mass-squared matrix  is transformed to the block
diagonal form by a separation of the massless Goldstone
boson fields,\lb $G^0=\cos\beta\,\chi_1+\sin\beta\,\chi_2$
and
$G^\pm=\cos\beta\,\varphi_1^\pm+\sin\beta\,\varphi_2^\pm$,
and the charged Higgs boson fields $H^\pm$ with mass
$M_{H^\pm}$,\\[-3mm]
 \bear{c}
H^\pm\!=\!-\!\sin\beta\,\varphi_1^\pm\!+\!\cos\beta\,\varphi_2^\pm
,\;\;\; M_{H^\pm}^2\!=\! \left[\nu\!-\!\dfrac{
\lambda_4+Re\, \lambda_5+Re\,
\lambda_{67}^+}{2}\right]v^2.\vspace{-2mm}
 \eear{Eq:mch}

\subsection{Neutral Higgs sector}

By definition $\eta_{1,2}$ are the standard $C$-- and $P$--
even (scalar) fields. The  field\\[-3mm]
\begin{equation}
A=-\sin\beta\,\chi_1+\cos\beta\,\chi_2\,,
\label{physfields}\vspace{-2mm}
 \end{equation}
is $C$--odd (which in the interactions with fermions
behaves as a $P$-- odd particle, i.e. a pseudoscalar). In
other words, the $\eta_{1,2}$ and $A$ are fields with
opposite CP parities (see e.g. \cite{Hunter} for details).

The decomposition (\ref{videf}) results in the symmetric
mass--squared matrix $\cal M$ in the $\eta_1$, $\eta_2$,
$A$ basis\\[-3mm]
 \bear{c}
{\cal M}=\begin{pmatrix}
M_{11} & M_{12} & M_{13} \\
M_{12} & M_{22} & M_{23} \\
M_{13} & M_{23} & M_{33}
\end{pmatrix},\\
 M_{11}=\left[c_\beta^2\,\lambda_1 +s_\beta^2\,\nu+
s_\beta^2 Re\,(\lambda_{67}^+/2+\lambda_{67}^-)\right]v^2,
 \\
M_{22}=\left[s_\beta^2\,\lambda_2+c_\beta^2\,\nu +c_\beta^2
Re\,(\lambda_{67}^+/2-{\lambda}_{67}^-)\right]v^2
, \\
M_{33}=\left[\nu-Re\,(\lambda_5-\lambda_{67}^+/2)
\right]v^2\equiv M_A^2,
 \\
M_{12}=-\left[\nu-\lambda_{345}
-Re\,3\lambda_{67}^+/2\right]c_\beta s_\beta v^2,
\\
M_{13}=-\left[\delta +Im{\lambda}_{67}^-/2\right]s_\beta
v^2,\;\; M_{23}=-\left[\delta -Im
{\lambda}_{67}^-/2\right]c_\beta v^2,\vspace{-2mm}
 \eear{Eq:M3by3}
where $c_\beta=\cos\beta$, $s_\beta=\sin\beta$. Note that
$M_{33}$ is equal to the mass squared of the CP--odd Higgs
boson  in the CP conserving case $M_A^2$.

The masses squared $M_i^2$ of the physical neutral states
$h_{1-3}$ are eigenvalues of the matrix $\cal M$. These
states are obtained from fields $\eta_1,\,\eta_2,\,A$ by a
unitary transformation $R$ which diagonalizes the  matrix
$\cal M$:\\[-3mm]
 \be
\begin{pmatrix}
h_1 \\ h_2 \\ h_3
\end{pmatrix}
= R
\begin{pmatrix}
\eta_1 \\ \eta_2 \\ A
\end{pmatrix},\qquad
     R=\begin{pmatrix}
R_{11} & R_{12} & R_{13} \\
R_{21} & R_{22} & R_{23} \\
R_{31} & R_{32} & R_{33}
\end{pmatrix}
\label{Eq:R33}\vspace{-2mm}
 \ee
with  $R{\cal M}R^T=diag(M_1^2,\,M_2^2,\,M_3^2)$. All
observable Higgs fields $h_i$, $H^\pm$, their masses and
couplings are  RPa independent, in contrast with original
fields $\phi_{1,2}$. The useful 2-step diagonalization
procedure is described in \cite{GK05}.

$\nabla$ With radiative corrections the physical states
$h_i$ become unstable, they have no asymptotic states. Mass
matrix   for these Higgs bosons become non-hermitian. This
effect can be neglected when the widths of Higgs bosons are
much smaller than the mass splitting. If one of masses
$M_i$ is close to another mass, a reasonable description of
the masses and couplings is given by an approximation in
which a (complex) matrix of polarization operators is added
to the mass matrix (\ref{Eq:M3by3}). Full treatment of this
problem demands a subtle theoretical analysis.

\subsection{Criterium for CP violation}

In general, the Higgs eigenstates $h_i$ (\ref{Eq:R33}) have
no definite CP parity since they are mixtures of fields
$\eta_{1,2}$ and $A$ having  opposite CP parities. Just
this mixing provides a CP nonconservation within the Higgs
sector since  the interaction of these Higgs bosons with
matter explicitly violates the CP--symmetry.

The eq.~\eqref{Eq:M3by3} shows that such mixing is absent
and the CP does not violated if and only if $
M_{13}=M_{23}=0$. The explicit form for these terms
\eqref{Eq:M3by3} shows that these two conditions can be
valid (at $\sin2\beta\neq 0$) if and only if
$\lambda_{67}^-$ and $m_{12}^2$ are real. In accordance
with \eqref{nudeldef} it means that the CP violation is
absent if all coefficients in potential of a real vacuum
form are real. Vice versa, the complexity of some
parameters of the potential in {\it a real vacuum form} is
a sufficient condition for CP violation in the Higgs
sector. Simple but cumbersome calculation shows that
similar conclusion is valid also for for $\sin2\beta=0$.
For an arbitrary form of Lagrangian  {\bf the necessary and
sufficient condition for CP violation in the Higgs sector}
can be written as complexity at least one
of combinations\\[-3mm]
 \be
\lambda_5^*(m_{12}^2)^2\,,\quad (\lambda_6^*+
\lambda_7^*)m_{12}^2\,,\quad \lambda_6^*\lambda_7.
 \label{newcond}\vspace{-2mm}
 \ee
Each this quantity is not RPa invariant one but these forms
are very simple. (For the  soft $Z_2$ violated potential
one should be $Im\,\lambda_5^*(m_{12}^2)^2\neq 0$ --
cf.~\cite{Froggatt:1992wt}). The RPa invariant conditions
for CP violation  \cite{invappr}, \cite{GH05} are more
complex.

$\nabla$ In MSSM, etc. CP symmetry can be violated by
interaction of Higgs fields with different scalar squarks,
etc. In this case the CP violated terms (like
$Im\,\lambda_{5-7}$, etc.) must be added in Lagrangian for
renormalizabilty.

\section{ Couplings  to gauge bosons and fermions}

Below we use  {\it in principle measurable relative
couplings} -- ratios of the couplings of each neutral Higgs
boson $h_i$  to the corresponding SM couplings
\\[-3mm]
 \begin{equation}
\label{Eq:gj} \chi_j^{(i)}=g_j^{(i)}/g_j^{\rm
SM}\,.\vspace{-2mm}
 \end{equation}
for  the gauge bosons $W$ or $Z$ and the quarks or leptons
($j=W,Z,u,d,\ell...$).

\bu The gauge bosons $V$ ($W$ and $Z$) couple only to the
CP--even fields $\eta_1$, $\eta_2$. For the physical Higgs
bosons $h_i$ (\ref{Eq:R33}) one obtains\\[-3mm]
 \be
\chi_V^{(i)}\!=\!\cos\beta\, R_{i1}\! +\!\sin\beta\, R_{i2}, \;\,
 V=W\mbox{ or } Z.
\label{Eq:W}
\end{equation}

\subsection{ Yukawa interaction}


The general form of Yukawa interaction couples 3-family
vector of the left-handed quark isodoublets $Q_L$ with
3-family vectors of the the right-handed field singlets
$d_R$ and $u_R$ and Higgs fields $\phi_i$. It allows large
FCNC effects and lead to  true hard violation of $Z_2$
symmetry via loop effect (see \cite{GK05} for details).

To have only the soft violation of $Z_2$ symmetry,  each
right-handed fermion should couple to only one scalar field, either
$\phi_1$ or $\phi_2$ \cite{Glashow:1977nt,Paschos:1976ay}.

\subsection{Model II}

We consider first most popular opportunity (realized also
in MSSM) named as Model II (cf. \cite{Hunter} for
classification). In this Model the physical reality allows
the description, where the fundamental scalar field
$\phi_1$ couples to $d$-type quarks and charged leptons
$\ell$, while $\phi_2$ couples to $u$-type quarks and this
interaction is diagonal (or
almost diagonal) in family index  $k$\\[-3mm]
 \be
  -\!{\cal L}_Y^{II}\!=\! \sum g_{dk}\bar{Q}_{Lk}
\phi_1 d_{Rk}\! +\! \sum g_{uk} \bar{Q}_{Lk} \tilde\phi_2
u_{Rk}\! +\! \sum g_{\ell k} \bar{\ell}_{Lk} \phi_1
\ell_{Rk} \!+\!{\rm h.c.}\vspace{-2mm}
 \label{YukII}
\end{equation}

The suitable choice of phases in RPh transformations makes
all Yukawa parameters real.  As it was written above,
different forms of Lagrangian can have different values of
$\tan\beta$. To underline that we use the mentioned
Lagrangian, we will supply (only in this section) quantity
$\beta$  by a subscript II, $\beta\to \beta_{II}$. (The RPa
transformation makes Model II property of Lagrangian hidden
and changes $\tan\beta$.)

The relative Yukawa couplings of the physical neutral Higgs
bosons $h_i$  (\ref{Eq:gj})  are identical for all
$u$--type and for all $d$--type quarks (and charged
leptons). They are expressed via elements of the
rotation matrix $R$ (\ref{Eq:R33}) as\\[-3mm]
 \be\label{Eq:chi-ud}
(MII):\qquad\chi_u^{(i)}= \dfrac{R_{i2}-i\cos\beta_{II}\,
R_{i3}}{\sin\beta_{II}},\;\;\; \chi^{(i)}_d
=\dfrac{R_{i1}-i\sin\beta_{II}\,R_{i3}}{\cos\beta_{II}}.\vspace{-2mm}
 \ee

In the cases of weak CP violating  and soft $Z_2$-violation
the relative coupling of the neutral scalar $h_i$  to the
charged Higgs boson is expressed via the couplings of this
neutral Higgs boson to the
gauge bosons and fermions:\\[-3mm]
 \bear{c}
\chi_{H^\pm}^{(i)}
=\left(1-\fr{M_i^2}{2M_{H^\pm}^2}\right)\chi_V^{(i)}
+\fr{M_i^2-\nu v^2}{2M_{H^\pm}^2}
Re\,(\chi_u^{(i)}+\chi_d^{(i)}).
 \end{array}\label{b2d3}
\end{equation}\vspace{-2mm}

$\nabla$ The  unitarity of the mixing matrix $R$ allows one
to obtain a number of {\bf relations between the relative
couplings of neutral Higgs particles}. These relations are
very useful in phenomenological analyzes.

First, the quantity $\tan\beta_{II}$  (coincideny with the
ratio $v_2/v_1$ only in a Model II form of Lagrangian) is
described via the basic couplings to $h_i$ as\\[-3mm]
\begin{equation}
\cot^2\beta_{II}\!= \!{\fr{(\chi_V^{(i)}\!-\!\chi_u^{(i)})}
{\left(\chi_d^{(i)}\!-\!\chi_V^{(i)}\right)^*}} \!=\!
{\fr{1\!-\!|\chi_u^{(i)}|^2}{|\chi_d^{(i)}|^2\!-\!1}}= \!
\fr{Im \chi_u^{(i)}}{Im \chi_d^{(i)}}\!=\!\sum_i
(Im\chi_u^{(i)})^2 .\vspace{-1mm} \label{Eq:tan-beta}
\end{equation}

1. {\em The pattern relation for each neutral Higgs
particle $h_i$} (for CP conserving case see
\cite{Ginzburg:2001ss,Ginzburg:2002wt}):\\[-2mm]
 \be\label{2hdmrel}
(\chi_u^{(i)} +\chi_d^{(i)})\chi_V^{(i)}=1+\chi_u^{(i)}
\chi_d^{(i)}\, .\vspace{-1mm}
\end{equation}

2. {\it A vertical sum rule for   {\bf all three} neutral
Higgs bosons
$h_i$} \cite{Grzadkowski:1999ye}:\\[-3mm]
 \be
\sum\limits_{i=1}^{3}(\chi_j^{(i)})^2=1\,\qquad
(j=V,d,u)\,. \label{vsr}\vspace{-2mm}
 \ee
For couplings to the gauge bosons this sum rule takes place
independently on a particular form of the Yukawa
interaction.

3.  {\em A horizontal  sum rule} \cite{Grzadkowski:1999ye}
for each neutral Higgs boson $h_i$:\\[-3mm]
\begin{equation}
|\chi_u^{(i)}|^2\sin^2\beta_{II}+|\chi_d^{(i)}|^2\cos^2\beta_{II}=1\,.\label{srules}
 \vspace{-2mm}\end{equation}
These sum rules guarantee that  the cross section 
of production of each neutral Higgs boson $h_i$  of the 2HDM by one
of types of quarks cannot be lower than that for the SM Higgs boson
with the same mass \cite{Grzadkowski:1999ye}.

4. Besides, the   {\it linear relation} follows directly
from Eqs.~(\ref{Eq:R33}), (\ref{Eq:chi-ud}):\\[-3mm]
 \bear{c}
\!\begin{array}{c}Re\left(\cos^2\beta_{II}
\chi_d^{(i)}\!+\!\sin^2\beta_{II}
 \chi_u^{(i)}\right)\!=\!\chi_V^{(i)}, \\[1mm]
 Im\left(\cos^2\beta_{II} \chi_d^{(i)}-\sin^2\beta_{II}
 \chi_u^{(i)}\right)=0.
\end{array}\vspace{-2mm} \end{array}\label{reimchi}
 \ee
5. {\it  The relation 
between CP violated parts of Yukawa couplings} is obtained by
exclusion of $\beta_{II}$ from
the equations \eqref{srules}, \eqref{reimchi}\\[-3mm]
  \be
 (1-|\chi_d^{(i)}|^2)\,Im\chi_u^{(i)}
  +(1-|\chi_u^{(i)}|^2)\,Im\chi_d^{(i)}=0\,.
  \label{newrelchi}\vspace{-2mm}  \ee

A number of applications of this set of relations is
discussed in \cite{GK05}.\\[-3mm]

$\nabla$ The observable quantities correspond the
Lagrangian  with  radiative corrections (RC). Then one can
treat the relations \eqref{2hdmrel}--\eqref{newrelchi} as
obtained from the renormalized parameters. For each
relative coupling (\ref{Eq:gj}) the RC are included in
both: the couplings of the 2HDM (in the numerator) and
those of SM (in the denominator). The largest RC to the
Yukawa $\phi \bar{q}q$ couplings are the one--loop QCD
corrections due to the gluon exchange. They are identical
in the SM and in the 2HDM and cancel in  ratios $\chi_u$
and $\chi_d$. The same is valid for purely QED  RC to all
basic couplings as well as for electroweak corrections
including virtual $Z$ or $W$ contributions.

The electroweak RC containing Higgs bosons in the loops are
different in the SM and 2HDM,  their values depend on the
parameters of 2HDM. These type of RC may modify slightly
some relations presented above. However, it is naturally to
expect that these RC are small ($\lesssim 1$~\%) except for
some small corners of parameter space.

\subsection{Model I}

We consider also, for completeness, this model, in which
all right handed fermions are coupled to one Higgs field
$\phi_1$.  The general RPa transformation makes this
property hidden, changing simultaneously $\tan\beta$. We
supply the parameter $\beta$ for the Model I form of
Lagrangian by subscript I.

The corresponding Model I Yukawa Lagrangian is similar to
that \eqref{YukII} with only change $\phi_2\to \phi_1$. For
this form of Lagrangian we have\\[-3mm]
 \be\label{Eq:chi-udMI}
 (M\;I):\qquad\chi_u^{(i)}=\chi_d^{(i)}\equiv \chi_f^{(i)}=
 \dfrac{[R_{i2}-i\cos\beta_{I}\,
R_{i3}]}{\sin\beta_{I}}.\vspace{-2mm}\ee

In this case only one of methods for finding of $\beta_I$
via observable quantities among series presented in
\eqref{Eq:tan-beta} works, $
\cot^2\beta_{I}=\sum_i\left(Im\chi_u^{(i)}\right)^2$, just
as vertical sum rules \eqref{vsr}. Other relations written
for Model II (\eqref{2hdmrel},
\eqref{srules}--\eqref{newrelchi} don't work for this
Model.

\section{Constraints for Higgs Lagrangian}

\subsection{Positivity (vacuum stability) constraints.}

To have a {\sl stable vacuum}, the potential must be
positive at large quasi--classical values of fields
$|\phi_k|$ ({\sl {positivity constraints}}) for an
arbitrary direction in the $(\phi_1,\phi_2)$ plane. These
constraints were obtained for the case of soft $Z_2$
violation (see e.g. \cite{dema}-\cite{Gunion:2002zf}), they
are\\[-3mm]
 \bear{c}
 \lambda_1>0\,, \quad \lambda_2>0,\;\,
\lambda_3+\sqrt{\lambda_1\lambda_2}>0,\;\;
\lambda_3+\lambda_4-|\lambda_5|+\sqrt{\lambda_1\lambda_2}>0.\end{array}
\label{positiv}
 \end{equation}

\subsection{ Minimum constraints}.

The condition for vacuum (\ref{Eq:min-cond}) describes the
{\it extremum} of potential but not obligatory the minimum.
The {\it minimum constraints} are  the conditions ensuring
that above extremum is a  minimum for all directions in
$(\phi_1,\phi_2)$ space, except of the Goldstone modes (the
physical fields provide the basis in the coset). This
condition is realized if the mass-matrix squared for the
physical fields is positively defined: $
M_{h_{1-3}}^2,\;M_{H^\pm}^2>0$.  Note that the change of
sign of $M_{H^\pm}^2$ \eqref{Eq:mch} with all positive
neutral mass squares correspond to the transition from
physical vacuum to the charged vacuum.

\subsection{Tree level unitarity constraints}.

The quartic terms of Higgs potential ($\lambda_i$)  lead,
in the tree approximation, to the s--wave Higgs-Higgs and
$W_LW_L$ and  $W_LH$, etc. scattering amplitudes for
different elastic channels. These amplitudes should not
overcome unitary limit for this partial wave  -- that is
{\it the tree-level unitarity constraint}.

The unitarity constraint  was obtained first
\cite{Glashow:1977nt} for the  minimal SM, with Higgs
potential $V= (\lambda/2)(\phi^\dagger\phi-v^2/2)^2$. Such
constraints for the 2HDM with a soft $Z_2$ violation and CP
conservation were derived in \cite{Akeroyd:2000wc}.

In the  general CP nonconserving case unitarity constraints
are written  in ref.~\cite{unitCP} as the bounds for the
eigenvalues  of the high energy Higgs--Higgs scattering
matrix $S_{Y\sigma}=16\pi\Lambda_{Y\sigma}$ for the
different quantum numbers of an initial state: total
hypercharge $Y$ and weak isospin $\sigma$ \eqref{scmat}.
(In each case left upper $2\times 2$ corner presents
scattering matrix for $Z_2$--even states and right--down
corner --- for $Z_2$--odd states, while coefficients
$\lambda_6$, $\lambda_7$ describe mixing among these
states.)

The unitarity constraint means that the eigenvalues of $S$
matrix are less than 1, therefore the eigenvalues of the
written matrices $\Lambda_{Y\sigma}$ limited as\\[-4mm]
 \be
 |\Lambda_{Y\sigma}|< 16\pi
\,.\label{unitconstr}\vspace{-8mm}
 \ee

 \bes\label{scmat}
 \bea
& \Lambda_{Y=2,\sigma=1}\!\!=\!\!
 \begin{pmatrix}
\lambda_1 & \lambda_{5}&\sqrt{2}\lambda_6\\
 \lambda_5^*&  \lambda_2&\sqrt{2}\lambda_7^*\\
\sqrt{2}\lambda_6^*&\sqrt{2}\lambda_7&
\lambda_3\!+\!\lambda_4
\end{pmatrix}\!\!,\;\;\;
 \Lambda_{Y=2,\sigma=0}\!\!=\!\;\lambda_3\!-\!\lambda_4,&\label{YS20}\\
 &
  \Lambda_{Y=0,\sigma=1}\!=\!\begin{pmatrix}
 \lambda_1 & \lambda_4&\lambda_6&\lambda_6^*\\
 \lambda_4&  \lambda_2&\lambda_7&\lambda_7^*\\
\lambda_6^*&\lambda_7^*&\lambda_3&\lambda_5^*\\
\lambda_6&\lambda_7&\lambda_5&\lambda_3
 \end{pmatrix},&\label{YS01}\\
 &
\Lambda_{Y=0,\sigma=0}=\begin{pmatrix}
3\lambda_1&2\lambda_3+\lambda_4&3\lambda_6&3\lambda_6^*\\
  2\lambda_3+\lambda_4& 3\lambda_2&3\lambda_7&3\lambda_7^*\\
 3\lambda_6^*&3\lambda_7^*&\lambda_3+2\lambda_4& 3\lambda_{5}^*\\
 3\lambda_6&3\lambda_7&
 3\lambda_5&\lambda_3+2\lambda_4\end{pmatrix}\,.&\label{YS00}
 \vspace{-4mm}
 \eea\ees

The eigenvalues of these matrices can be found as  roots of
equations of the 3-rd or 4-th degree. It is useful to start
diagonalization from corners of these matrices,
corresponding to fixed values of the $Z_2$ parity. This
particular diagonalization transform $\Lambda_{Y\sigma}$ to
the form with diagonal elements  coincident with
eigenvalues found in \cite{Akeroyd:2000wc} (for soft $Z_2$
violation without CP violation) with sole change
$\lambda_5\to |\lambda_5|$.

Let us remind that {\bf\bm all diagonal matrix elements
$M_{ii}$ of a Hermitian matrix $||M_{ij}||$ with maximal
and minimal eigenvalues $\Lambda_\pm$  lie between them,
$\Lambda_+\ge M_{ii}\ge \Lambda_-$.} It means that the
mentioned corrected constraints from \cite{Akeroyd:2000wc}
form \underline{necessary} conditions for unitarity. These
constraints are enhanced due to terms describing hard $Z_2$
violation.

\section{2HDM and observations}

Some possible observation will be clear signal in favor of
difference of our world from that described by minimal SM
and in the attempt to check whether EWSB is given by 2HDM.
{\it(i)} If more than one Higgs boson will be observed.
{\it(ii)} If -- in the case of observation of single Higgs
boson -- the strong difference in the couplings of Higgs
boson with matter from SM predictions will be observed.

$\Box$ The most difficult for analysis is the case of
realization of {\bf SM-like physical picture}: the lightest
Higgs boson $h_1$ is  similar to the Higgs boson of the SM
while other Higgs bosons escape observation being too heavy
(or weakly coupled with matter) \cite{Ginzburg:2001ss}.

$\nabla$ {\bf Heavy Higgs bosons in 2HDM}. Besides, many
authors assume {\it in addition} that masses of other Higgs
bosons $M$ are close to the scale of new physics, $M\sim
\Lambda$, and that the theory should possess an explicit
{\bf decoupling property}, i.e. {\it the correct
description of the observable phenomena must be valid for
the (unphysical) limit $M \to\infty$}
\cite{Kanemura:1996eq, Gunion:2002zf}. (This property --
independence from phenomena at $p>\Lambda$ -- is necessary
feature of any consistent theory describing phenomena at
$p\ll \Lambda$ but only if limit $\Lambda\to\infty$ has
physical sense \cite{Appelquist:tg}.) The 2HDM allows also
for another realization of the SM--like physical picture.

Looking for mass matrix \eqref{Eq:M3by3} one can see that
the large masses of Higgs particles may arise from large
parameters $\nu$ or $\lambda's$, or both. Obviously, large
values of $\lambda$'s may be in conflict with unitarity
constraints, which is not the case for large $\nu$. The
case $\nu\gg |\lambda_i|$ correspond to a decoupling
regime, while the case of small $\nu$ and not very high
$|\lambda_i|$ allows quite another realization of SM--like
scenario. Both these opportunities were analyzed in detail
in \cite{GK05}.

This analysis allows one to show that {\bf the natural set
of parameters of  2HDM} correspond to the case of soft
violation of $Z_2$ symmetry and $|\nu|, |\lambda_5|\ll
|\lambda_{1-4}|$.  From this point of view {\it the
decoupling case of 2HDM with $\nu\gg |\lambda_i|$ is
unnatural}.

The 2HDM with natural set of parameters (not in the
decoupling case) and SM can be distinguished via
observation of Higgs boson production at Photon Collider
\cite{Ginzburg:1999fb}.

\section{Possible relation to a history of
time}

The modern description of the beginning of time  contains
assumption about SM Higgs mechanism of EWSB. In the hot
primitive medium after Big Bang the effective Higgs
potential of SM is added by term $cT^2\phi^2/2$. It changes
standard mass term  $-m^2\phi^2/2$ so that the v.e.v. of
Higgs field \eqref{SMvac} with growth of temperature
decreases as $\sqrt{m^2-cT^2}/\sqrt{2\lambda}$. At the
temperature $T_c\approx m/\sqrt{c}$ (determined with
accuracy to quantum corrections) we have phase transition.
After Big Bang, when  $T>T_c$, we had $\la\phi\ra=0$, EWSB
was not broken, particles were massless, providing
exponential inflatory expansion of Universe. After cooling
to $T<T_c$ we come to our world with massive particles,
etc. and nonzero vacuum energy $E_{vac}$ -- see for details
references in \cite{PDG}.

One can imagine two opportunities for this picture.

\bu First, the inflation mechanism is given by Higgs field,
responsible for EWSB. In this case,  possible existence of
two vacua in 2HDM opens new opportunity in the history of
time. Here in the hot medium the effective potential
\eqref{Eq:Lagr-Higgs} is added by
terms\\[-5mm]
 \be
\left[c_{11}(\phi_1^\dagger
\phi_1)\!+\!c_{12}(\phi_1^\dagger
\phi_2)\!+\!c_{12}^*(\phi_2^\dagger
\phi_1)\!+\!c_{22}(\phi_2^\dagger
\phi_1)\right]\fr{T^2}{2}\,.\label{tempeff}\vspace{-2mm}
 \ee
They change mass terms of our Lagrangian so that
immediately after Big Bang the Universe expands inflatory
in the same manner as in minimal SM. The subsequent fate of
Universe depends on values of parameters.

In one case at the growth of time the EWSB vacuum
$\la\phi_1\ra=\la\phi_2\ra=0$ at some critical temperature
is transformed directly to the neutral vacuum
\eqref{phvac2}. In this case transformation of Universe are
completely the same as those discussed in respect of
minimal SM.

In the other case at the growth of time the EWSB vacuum
$\la\phi_1\ra=\la\phi_2\ra=0$ at some critical temperature
$T_{c1}$ is transformed first in the charged vacuum
\eqref{chvac} and only with subsequent growth of time at
some  temperature $T_{c2}<T_{c1}$ the charged vacuum is
transformed into well known neutral vacuum \eqref{phvac2}.
The life of Universe in the period when $T_{c2}<T<T_{c1}$
can be quite unusual. In this stage the medium is
absolutely non-transparent for light (photon is massive),
the transformations of particles are quite different from
modern, the $C$ violation for particles (vacuum is charged)
can leave after second phase transition an Universe with
residual CP violation and influence for baryon asymmetry,
etc. Besides, some small domains of charged phase appeared
from fluctuations in one of phase transitions can leave up
to our time, influencing for modern observations. Some of
these opportunities can be excluded quickly by first
analysis, but the other must be studied in future in
detail.

\bu Second, inflation can be related to a specific {\it
inflanton} Higgs field $\phi_0$ with varying in time v.e.v.
$\la\phi_0\ra=U_0(t)$. This field should interact with
Higgs field responsible for EWSB like \eqref{tempeff}, the
effective Higgs potential is added by term\vspace{-4mm}
 \be
\left[a_{11}(\phi_1^\dagger
\phi_1)\!+\!a_{12}(\phi_1^\dagger
\phi_2)\!+\!a_{12}^*(\phi_2^\dagger
\phi_1)\!+\!a_{22}(\phi_2^\dagger
\phi_1)\right]\fr{U_0^2}{2},\label{tempeff1}\vspace{-2mm}
 \ee
where coefficients $a_{ij}$ can be both positive and
negative. Therefore, during inflation effective mass term
of the EWSB Higgs field varies with time as $m_{ij}^2\to
m_{ij}^2-c_{ij}T^2- a_{ij}U_0^2(t)$. It can results in even
more complex sequence of phase transitions than that
discussed above (e.g., with restoration of $SU(2)\times
U(1)$ symmetry in some intermediate period).

Both these opportunities should be analyzed in
future.\\[-3mm]

{\bf Acknowledgments.} I am thankful M. Krawczyk, and I.
Ivanov for very  fruitful collaboration. Many results
presented here were obtained together with them. This
research has been supported by Russian grants RFBR
05-02-16211, NSh-2339.2003.2.

\end{document}